\documentclass[a4paper,11pt]{article}
\usepackage{pos}

\newcommand{\bibi}[8]{\bibitem{#1} #2, \emph{#3}, #4, \href{https://doi.org/#5}{\emph{#6} \textbf{#7} #8}}

\newcommand{\pd}[1]{\partial\,#1}
\newcommand{\E}[1]{\times 10^{#1}}

\title{The imprint of protons on the emission of extended blazar jets}
 \ShortTitle{ExHaLe-jet}

\author*[a,b]{Michael Zacharias}
\author[c]{Anita Reimer}
\author[a]{Andreas Zech}

\affiliation[a]{Laboratoire Univers et Théories, Observatoire de Paris, Université PSL, CNRS, Université de Paris, \\ 5 pl Jules Janssen, 92190 Meudon, France}

\affiliation[b]{Centre for Space Science, North-West University, \\ Potchefstroom, 2520, South Africa}

\affiliation[c]{Institut f\"ur Astro- und Teilchenphysik, Leopold-Franzens-Universit\"at Innsbruck, \\ A-6020 Innsbruck, Austria}

\emailAdd{michael.zacharias@obspm.fr}
\emailAdd{mzacharias.phys@gmail.com}

\abstract{Blazars – active galaxies with the jet pointing at Earth – emit across all electromagnetic wavelengths. The so-called one-zone model has described well both quiescent and flaring states, however it cannot explain the radio emission. In order to self-consistently describe the entire electromagnetic spectrum, extended jet models are necessary. Notably, kinetic descriptions of extended jets can provide the temporal and spatial evolution of the particle species and the full electromagnetic output. Here, we present the initial results of a recently developed hadronic extended-jet code. As protons take much longer than electrons to lose their energy, they can transport energy over much larger distances than electrons and are therefore essential for the energy transport in the jet. Furthermore, protons can inject additional leptons through pion and Bethe-Heitler pair production, which can explain a dominant leptonic radiation signal while still producing neutrinos. We will present a detailed parameter study and provide insights into the different blazar sub-classes.
}

\FullConference{37$^{\rm{th}}$ International Cosmic Ray Conference (ICRC 2021)\\
		July 12th -- 23rd, 2021\\
		Online -- Berlin, Germany}

\begin{document}
\maketitle

\section{Introduction}
As a type of radio-loud active galactic nuclei, blazars harbor a relativistic jet pointing at Earth \cite{up95}. Due to relativistic beaming effects, the jet radiation is significantly enhanced allowing for detailed studies of the jet physics. The multiwavelength spectral energy distribution (SED) shows two broad components. The low-energy one peaks in the infrared to X-ray domain and is attributed to synchrotron emission of relativistic electrons. These same electrons may also be responsible for the high-energy component peaking in the $\gamma$-ray domain through inverse-Compton scattering of ambient photon fields, such as the self-made synchrotron photons (synchrotron-self Compton, SSC) or external photons from the accretion disk (AD), the broad-line region (BLR) and/or the dusty torus (DT). If relativistic protons are present in the jet, they can contribute to the $\gamma$-ray component through synchrotron emission, or through proton-photon interactions producing a cascade of secondary electrons and positrons \cite{bea13,c20}.

In most cases the SED is well reproduced using the one-zone emission model, where a single, spherical, and dense region is responsible for the radiative output. Especially during flaring events -- and blazars exhibit lots of flares, e.g. \cite{zea17} -- the light-travel time restricts the emission region to less than a parsec (at times much less than that). This is orders of magnitudes smaller than the length of jets, which is hundreds of kilo-parsecs. However, the relatively high particle density in the one-zone emission region result in a high opacity at radio frequencies. Furthermore, radio VLBI observations resolving jet structures on parsec scales (and sometimes below that \cite{eht19}) reveal the presence of relativistic particles also on larger scales beyond the typical location of the one-zone emission region.

This already indicates that the one-zone model is not a good representation of the jet as a whole. Additionally, in several cases the one-zone model also fails at other frequency bands to reproduce the SED. One example is the blazar AP Librae, where the very-high-energy (VHE) $\gamma$-ray emission cannot be reproduced through the one-zone model, and parts of the extended jet are required \cite{hea15,zw16}. Similarly, the recent detection of extended VHE $\gamma$-ray emission in the radio galaxy Centaurus~A \cite{Hea20} -- along with its peculiar spectrum \cite{Hea18} -- warrant multi-zone or extended models.

In recent years, several leptonic, extended jet models have been developed in order to model the emission from blazars -- e.g., \cite{pc13,lea19}. However, the potential association between a blazar and an astrophysical neutrino \cite{ice18} has increased the interest of hadronic emission scenarios in the blazar environment. Here, we present the novel code \textit{ExHaLe-jet} \cite{zea21}, which calculates the emission of extended blazar jets employing a hadro-leptonic scenario.

\section{ExHaLe-jet}
\begin{figure}[th]
\centering 
\includegraphics[width=0.65\textwidth]{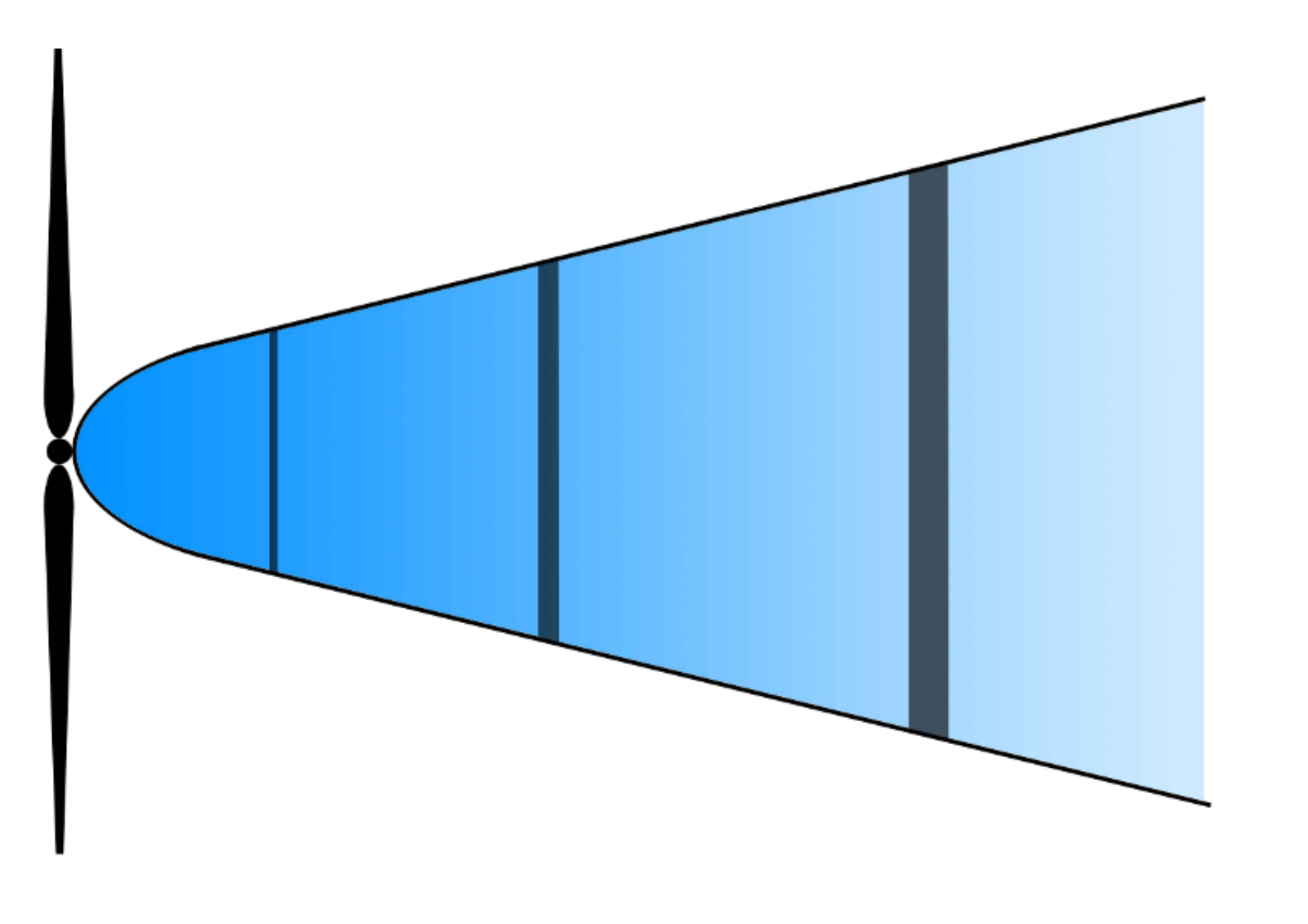}
\caption{Sketch of the model showing the extended jet, the decreasing plasma density (blue shades) and exemplary slices (dark). At small distances below $z_{acc}$, the geometry is parabolic, while it is conical at larger distances.
Figure courtesy of Jonathan Heil.
}
\label{fig:sketch}
\end{figure} 
In order to describe blazar jets, the code solves the Fokker-Planck equation describing the evolution of the particle distributions within a fixed jet geometry, as well as a fixed evolution of the bulk Lorentz factor $\Gamma_b$. Following \cite{lea19}, we assume an initial parabolic geometry, where the bulk flow accelerates, followed by a conical section with a constant bulk speed. The radius $R$ is proportional to the jet opening angle, which is a function of $\Gamma_b$. Specifically,

\begin{align}
	\Gamma_b (z) &= \begin{cases}
        1.09 + (\Gamma_{b,{\rm max}}-1.09) \frac{\sqrt{z}-\sqrt{z_0}}{\sqrt{z_{\rm acc}}-\sqrt{z_0}} & z\leq z_{\rm acc} \\
        \Gamma_{b,{\rm max}} & z>z_{\rm acc}
    \end{cases}
    \label{eq:BLF} \\
    R(z) &= 10 z_0 + (z-z_0) \tan{(0.26/\Gamma_b(z))}
    \label{eq:jetradius},
\end{align}
where $\Gamma_{b,{\rm max}}=30$ is the maximum bulk Lorentz factor, $z$ is the distance from the black hole, $z_0$ the minimum distance of the jet from the black hole (set at the innermost stable circular orbit), and $z_{\rm acc}=1\,$pc is the length of the (bulk) acceleration region.

At the base of the jet, a primary distribution of protons and electrons is injected, which is self-consistently evolved along the jet flow. In order to calculate the steady-state particle distributions $n_i$ at any given $z$, the jet is cut into numerous slices -- c.f., Fig.~\ref{fig:sketch} -- wherein the kinetic equation for every charged particle species (protons, charged pions, muons, electrons\footnote{We include both electrons and positrons in the term ``electron''.}) is solved:

\begin{align}
     \frac{\pd{n_i(\chi,t)}}{\pd{t}} = \frac{\pd{}}{\pd{\chi}} \left[ \frac{\chi^2}{(a+2)t_{\rm acc}} \frac{\pd{n_i(\chi,t)}}{\pd{\chi}} \right]
	 - \frac{\pd{}}{\pd{\chi}} \left( \dot{\chi}_i n_i(\chi,t) \right) + Q_i(\chi)
	 - \frac{n_i(\chi,t)}{t_{\rm esc}} - \frac{n_i(\chi, t)}{\gamma t^{\prime}_{i,{\rm decay}}} 
	 \label{eq:fpgen}.
\end{align}
This equation is given as a function of normalized particle momentum $\chi=\gamma\beta$. It contains acceleration terms of the Fermi-I and II type (parameterized through the energy-independent acceleration time scale $t_{\rm acc}$), radiative and adiabatic cooling terms $\dot{\chi}$, injection\footnote{The spectral shape of the primary injection is the same in each slice, while the normalization depends on the slice-parameters}, escape (parameterized through $t_{\rm esc}$) and decay (given by the proper decay time $t^{\prime}_{i,{\rm decay}}$), if applicable. 

The interaction of protons with the ambient radiation fields (both internal and external) produce pions. While the neutral pions decay quasi-instantaneously into photons, the charged pions decay into muons, which then decay into electrons. Additionally, pairs are created through Bethe-Heitler- and $\gamma$-$\gamma$ pair production. These secondary pairs are transported downstream in the jet decreasing the ratio of the proton-to-electron density with increasing distance $z$ from the black hole. We include synchrotron cooling for all (charged) species, while the electrons also inverse-Compton scatter the ambient photon fields.

We set an initial magnetic field of $B(z_0)=200\,$G, which is then self-consistently evolved through the relativistic Bernoulli equation \cite{zea15}. Further details can be found in a forthcoming publication \cite{zea21}.

\section{First results}
\begin{figure}[th]
\centering 
\includegraphics[width=0.85\textwidth]{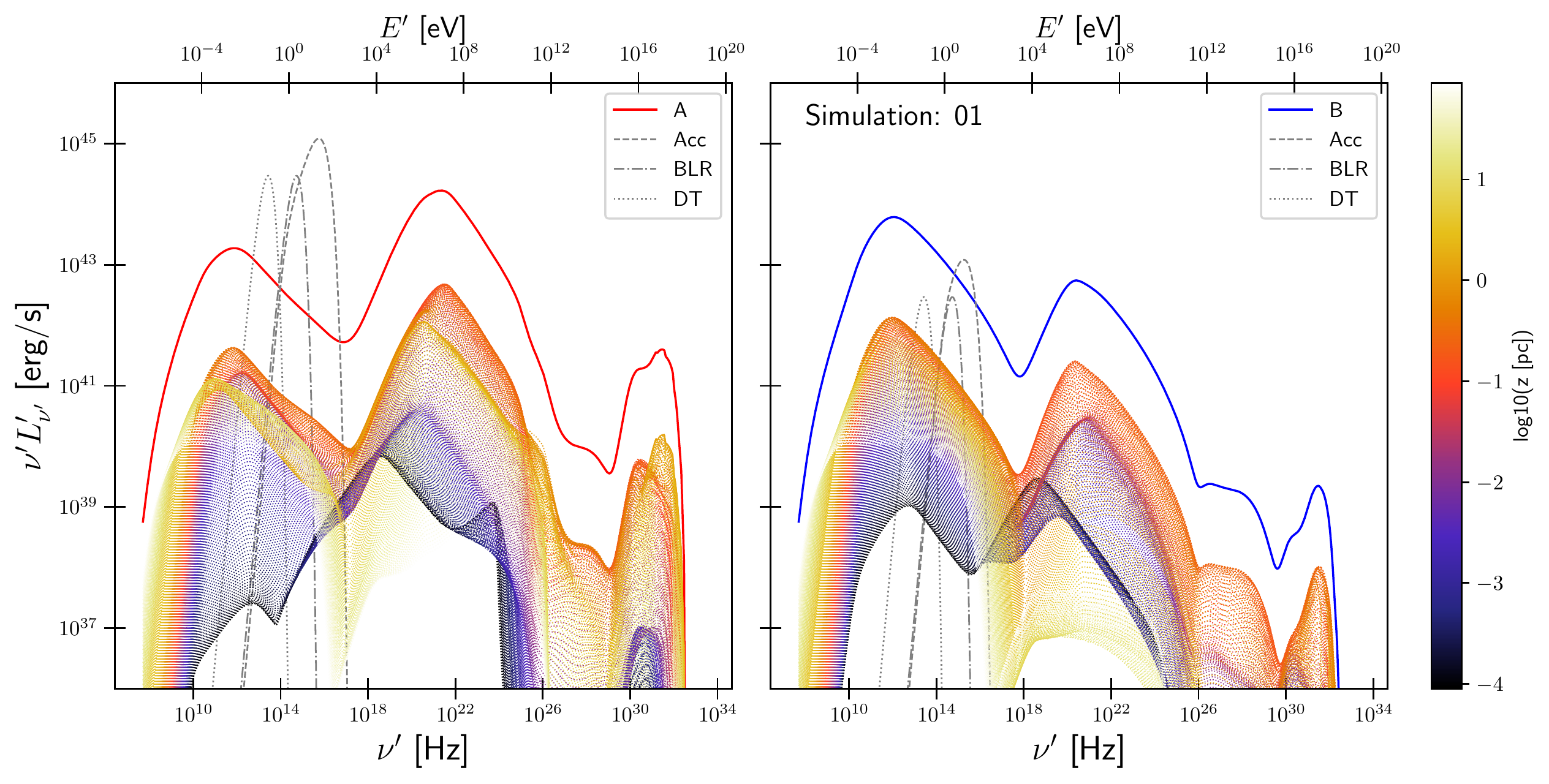}
\caption{Evolution of the spectral luminosity in the observer's frame with distance $z$ along the jet (color code). The thick lines mark the total spectra, while the gray lines mark the external fields as indicated. In this figure, we do not consider the absorption of $\gamma$ rays in the external fields \textit{outside} of the jet.
}
\label{fig:sim01_dist}
\end{figure} 
\begin{figure}[th]
\centering 
\includegraphics[width=0.85\textwidth]{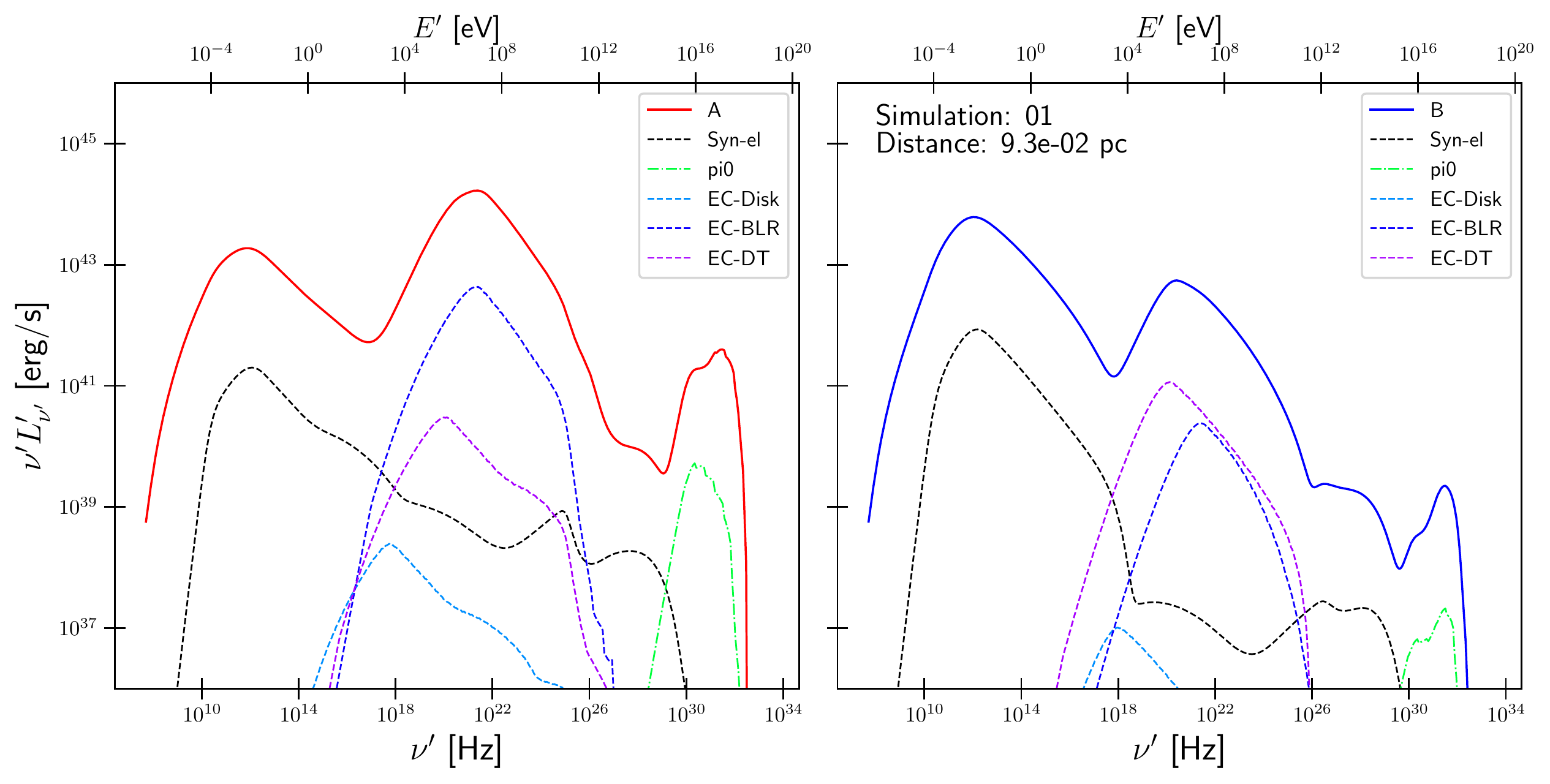}
\caption{Total spectra (thick lines) as in Fig.~\ref{fig:sim01_dist}, while the other lines show different spectral components as indicated at a distance of $z\sim 0.1\,$pc.
}
\label{fig:sim01_comp}
\end{figure} 
\begin{table}
\caption{Overview of the model parameters. The value of the AD Eddington ratio defines the cases A and B shown in Figs.~\ref{fig:sim01_dist} and \ref{fig:sim01_comp}.}
\begin{tabular}{lcl|c}
Definition		& \multicolumn{2}{c|}{Symbol} 		& Value  \\
\hline
Disk Eddington ratio    & $l_{\rm edd}$  &   & $10^{-1}$ \textcolor{red}{(A)},  $10^{-3}$ \textcolor{blue}{(B)}  \\
BLR temperature    & $T_{\rm BLR}$  & [K]  & $10^4$   \\
DT temperature    & $T_{\rm DT}$  & [K]  & $5\E{2}$   \\
Jet length    & $z_{\rm term}$  & [cm]  & $3.09\E{20}$   \\
Length of acceleration region    & $z_{\rm acc}$  & [cm]  & $3.09\E{18}$  \\
Maximum bulk Lorentz factor    & $\Gamma_{b,{\rm max}}$  &   & $30$  \\
Jet viewing angle    & $\theta_{\rm obs}$  & [rad]  & $3.33\E{-2}$  \\
Initial magnetic field strength    & $B(z_0)$  & [G]  & $200$  \\
Minimum proton Lorentz factor    & $\gamma_{p,1}$  &   & $2$   \\
Maximum proton Lorentz factor    & $\gamma_{p,2}$  &   & $2\E{8}$   \\
Proton spectral index    & $p_p$  &   & $2.5$   \\
Minimum electron Lorentz factor    & $\gamma_{e,1}$  &   & $1\E{2}$   \\
Maximum electron Lorentz factor    & $\gamma_{e,2}$  &   & $1\E{5}$  \\
Electron spectral index    & $p_e$  &   & $3.0$  \\
\end{tabular}
\label{tab:freepara}
\end{table}
In both Figs.~\ref{fig:sim01_dist} and \ref{fig:sim01_comp} we present two simulations (A and B) using \textit{ExHaLe-jet} with the model parameters given in Tab.~\ref{tab:freepara}. The two simulations are distinguished merely by the luminosity of the external fields. Simulation A (left panels) has a bright accretion disk radiating at $10\%$ of the Eddington luminosity, while in simulation B (right panels) the disk luminosity is 2 orders of magnitude less. As we assume that the BLR and DT luminosities depend on the accretion disk luminosity (re-radiating $10\%$ of the disk), their luminosities are also decreased in simulation B. 

As can be seen in Fig.~\ref{fig:sim01_dist}, the SED in simulation A is dominated by the $\gamma$-ray component, while in simulation B it is the low-energy component. Figure~\ref{fig:sim01_dist} shows the evolution of the emission with distance (color code). In fact, the low-energy component is relatively evenly emitted between $0.01\,$pc and $10\,$pc from the black hole, while the high-energy component mostly originates between $0.1\,$pc and $1\,$pc. This is true for both simulations A and B. Figure~\ref{fig:sim01_comp} shows the radiative components at a distance of $0.1\,$pc from the black hole. Leptonic radiation processes dominate most parts of the spectrum. The low-energy component is -- as expected -- electron synchrotron emission, while the high-energy component is dominated by inverse-Compton emission scattering the BLR (case A) and a mix of the BLR and the DT (case B).

However, we also notice a third component at frequencies beyond $10^{30}\,$Hz, which stems from the decay of neutral pions. One can see that its shape, distance-evolution, and overall flux strongly depends on the external fields. Moreover, between the high-energy component and the pion bump is a ``bridge'' of electron-synchrotron emission. This is produced from highly-energetic secondary pairs. These features indicate that protons do play an important role. While they might not show up directly in the emission spectrum, they are a vital source for highly energetic pairs. 
Unfortunately, owing to interactions with the EBL and the CMB, the pion bump is not observable at Earth.

\section{Summary}
In this presentation, we have shown the first result employing a newly-deveolped, extended hadro-leptonic jet code -- \textit{ExHaLe-jet}. It describes the evolution of both protons and electrons, as well as the secondary particles along the flow of a blazar jet. Injecting a primary proton and electron distribution, as well as an initial value of the magnetic field at the jet base, these quantities are evolved self-consistently along the jet. The bulk flow evolution and the jet geometry are fixed.

With the parameter set given in Tab.~\ref{tab:freepara}, we obtain the classical, double-humped SED structure. However, the total SED contains radiation emitted over significant parts of the jet with the main contributions between $0.01\,$pc and a few parsec from the black hole. These are much larger scales than in the typical one-zone model. In our setup, the SED is dominated by electron-synchrotron and inverse-Compton emission on the external photon fields. In fact, the latter have a strong impact on the final SED, as the case with strong external fields shows a Compton dominance larger than unity, while for weak external fields the opposite is true.

While protons do not contribute directly to the emission in this example, they are a vital contributor to highly energetic secondaries. These secondaries are important for the radiation processes and may explain the observed ratio of electrons to protons \cite{snm20} in jet lobes. The protons are also responsible for the production of neutrinos, which may be observed at Earth. Lastly, we would like to note that a different parameter set may increase proton-synchrotron emission reducing the leptonic dominance in the SED. For further discussions and details, we refer to \cite{zea21}.


%
%
%

\end{document}